# BubbleID: A Deep Learning Framework for Bubble Interface Dynamics Analysis

C. Dunlap,[1)] C. Li,[1)] H. Pandey,[1)] N. Le,[2)] and H. Hu[1, a)]

[1]*Department of Mechanical Engineering, University of Arkansas, Fayetteville, Arkansas, 72701, USA*

[2]*Department of Electrical Engineering and Computer Science, University of Arkansas, Fayetteville, Arkansas, 72701,*

*USA*

This paper presents BubbleID, a sophisticated deep learning architecture designed to comprehensively identify both static and dynamic attributes of bubbles within sequences of boiling images. By amalgamating segmentation powered by Mask R-CNN with SORT-based tracking techniques, the framework is capable of analyzing each bubble's location, dimensions, interface shape, and velocity over its lifetime, and capturing dynamic events such as bubble departure. BubbleID is trained and tested on boiling images across diverse heater surfaces and operational settings. This paper also offers a comparative analysis of bubble interface dynamics prior to and post-critical heat flux (CHF) conditions.

**I. INTRODUCTION**

The rapid growth of high-power applications necessitates more efficient cooling systems. Thermal management is becoming a bottleneck in many growing industries (e.g., data centers [1], nuclear power plants, high power density electronics, the power grid, electric vehicles [2], etc). Since traditional single-phase cooling methods are not able to sufficiently meet this demand, more sophisticated two-phase methods are being explored and, in some cases, implemented as alternatives. By leveraging the high latent heat, two-phase cooling can offer more efficient heat dissipation. In particular, pool boiling and flow boiling methods are promising avenues for cooling. However, several complexities arise when implementing boiling methods such as instabilities (e.g., critical heat flux, flow reversal [3]), complexities in setup, and a lack of physical understanding of the boiling phenomena. The nucleate pool boiling regime is the ideal regime for immersion cooling to operate in as it offers a high heat flux with relatively low superheats [4]. Beyond the nucleate regime is the transition boiling regime and the point of this switch, or the critical heat flux (CHF), is a major problem for physical implementations. At this point, a vapor layer begins to cover the heating surface and acts as an insulator resulting in rapid deterioration of the heat transfer and a subsequent rapid increase of temperature [5]. This can be detrimental to the system by leading to overheating or burnout. To avoid this, current implementations, operate with a high factor of safety so the full benefit of the nucleate regime pool boiling is not realized. Another limiting factor for implementations is that boiling is a complex phenomenon that is not fully understood. A better understanding of the boiling phenomenon is needed to design for improving the performance of boiling applications while

---

[a)] Corresponding Author: Han Hu (hanhu@uark.edu)

maintaining safety and efficiency. Many groups have performed experimental correlations, modeling, image analysis, etc. to improve this understanding.

Boiling image datasets contain a substantial amount of information and have been used extensively to glean insight into the boiling process through boiling regime classification, correlation development, bubble characteristic extraction, etc. The analysis of such image datasets, however, is made difficult due to their large size. To capture the change in the bubbles, high-speed cameras are typically required which produce thousands of images each second. With such large datasets, computer processing methods are needed to analyze the data. Traditional image processing methods involved computationally large codes for analyzing data [6]. Manual image processing has been used for identifying boiling characteristics such as departure frequency, diameter, and bubble velocity [7]. Other image processing methods have been used for approximating bubble parameters such as contact angles [8], or bubble growth rate [9]. Oikonomidou et al used four different image-processing algorithms developed by different universities for analyzing bubbles in the absence of gravity [10]. Some methods used MATLAB image processing to determine contact line diameter, bubble height, bubble volume, equivalent diameter, and contact angles. Sadaghiani et al sought to explore the effects of bubble coalescence through the use of experimental pool boiling data with different prepared surfaces and image processing [11]. They reported the bubble nucleation and growth rates, bubble departure diameter, and frequency and did an in-depth analysis of coalescence. Villegas et al used image processing to distinguish bubbles from a background in grayscale images to determine size, and shape and follow their position through successive images to determine trajectory to measure velocity [12]. The advancement of computer vision methods with machine learning has addressed some of the limiting issues in these traditional methods. They have aided the time and accuracy of such analysis and have introduced new forms of available analysis. These data-driven models have found applications with furthering understanding and improving boiling research and have been used for classifying boiling regimes, generating new physical descriptors [13], heat flux prediction, and extracting bubble statistics.

As a typical data-driving method, segmentation models have been largely used in image analysis for bubbles and two-phase flow for several end goals. Torisaki and Miwa used a Yolo v3 backbone for semantic segmentation of images from a gas-liquid two-phase flow [14]. From this, they extracted void fraction, approximate equivalent diameter, bubble aspect ratio, etc. Instance segmentation models allow for distinction between individual bubbles to be made and can enable the ability to gather more advanced information. Zhang et al used Mask R-CNN models and transfer learning to aid image processing for lab-on-a-chip data [15]. They trained each droplet or bubble with an annotated ground truth bounding box and mask. For boiling specifically, semantic and instance segmentation models have been used for vapor fraction prediction and bubble identification [16]. Cui et al also used Mask R-CNN along with ResNet101 for identifying bubbles in bubbly flow [17]. Seong et al used U-Net to identify bubbles in flow boiling images and determine if each bubble was coalesced, condensing, sliding, growing, or nucleation [18]. Soibam et al used a CNN model to generate masks for boiling images [19]. They used these masks to track the bubbles and determine coalescences, nucleation rate, and oscillation. These present implementations of segmentation models on boiling image data were used primarily for single image analysis to obtain general features such as bubble diameters, count, and vapor fraction. Just focusing on a single image analysis can make it difficult to fully realize what is happening in a single frame. For example, by just looking at an image it can be difficult to determine if coalescence is occurring or if the bubbles are overlapping while the inclusion of video frames can remove the ambiguity.

Motivated by the desire to understand the dynamic nature captured through consecutive boiling video frames, temporal models have begun to be explored. Tracking and multi-object tracking (MOT) models allow for detecting the same object across frames at different points in time. Multi-object tracking methods are growing in application and sophistication. These models have found applications in security, car safety, and driverless cars, and have also led to significant improvement in boiling image analysis applications. These models are constantly improving to account for issues commonly seen in early iterations such as occlusions, deforming objects, scaling, etc. The application of these models to boiling data has allowed for more advanced and efficient analysis than previously achieved. However, there are several issues encountered when trying to apply these MOT methods to boiling images [20]. For example, two or more bubbles can merge, and bubbles can be covered by another, the shape of bubbles is unpredictable and varies as they depend on pressure and heat flux. Also, reflections and shadows from other bubbles can increase the noise and difficulty of maintaining consistent tracking. Suh et al developed a framework titled Vision-IT for analyzing images from boiling and condensation [21]. It uses Mask-RCNN for detection and a tracking model which uses the Crocker-Grier algorithm. To help improve the tracking, they took advantage of the typical vertical and lateral movements of boiling and added weight to the x-coordinate feature. This Vision-IT framework has been used for several applications by their group [22,23]. Through the utilization of the Vision-IT framework, Chang et al identified features from microgravity flow boiling [22]. They filtered out bubbles with switching IDs and the bubbles that were cut out of frame for their analysis. From this work, they gathered different bubble statistics (i.e., bubble count, size, aspect ratio), wetting front, interfacial length, wavelength, vapor layer thickness, vapor fraction, bubble velocity, etc.

Despite their proven success, the majority of machine learning models on boiling video analysis are limited to acquiring static spatial features from single images, which neglects to account for the dynamic nature of boiling. In this work, we present a framework, BubbleID, for capturing both static and dynamic features of individual bubbles and the overall pool present in boiling image sequences. Within the framework, we define a process for acquiring individual bubble characteristics such as each bubble's size and morphology. We also propose a method of approximating a novel feature, the bubble interface velocity. The framework can also be used to extract dynamic features (e.g., departure rate) and static global features (e.g., vapor fraction or bubble count). It is also used to distinguish between vapor fractions accounting for all bubbles and only attached bubbles in a single frame. The rest of the paper is organized as follows. Section II detailed the experiment setup, proposed BubbleID framework, and analysis metrics. Section III provides the experiment results and its discussion. The conclusion of this paper is given in section IV.

## II. EXPERIMENT AND METHODOLOGY

In this section, the data preparation experiment, machine learning models, and the analysis methodology of the build model outputs for bubble dynamics are presented as follows.

### A. Data Preparation

High-speed images during pool boiling experiments were used for training and testing machine learning models. These models include instance segmentation models with different amounts of classes. These models were all trained and tested on in-house experimental data which span multiple experiments. The following sections describe the specifics of the pool boiling experiments and how the images and labels were generated.

Multiple boiling datasets were used for the training and utilization of the machine learning models presented in the paper, as summarized in Table I. These data sets include the authors' past steady-state boiling tests on a variety of surfaces, including polished copper [24], copper foams [24], and copper microchannels [25,] and newly performed transient boiling tests. A detailed description of the pool boiling facility can be found in Ref [24]. For the experiments, nine cartridge heaters (Omega Engineering HDC19102) were inserted into the base of the block for heating and powered by a DC power supply (Magna-Power SL200-7.5). Four thermocouples (Omega Engineering TJ36-CPSS-032U-6) were mounted equidistance (0.1 in) from each other in the neck of the copper block for approximating the heat flux at the surface via regression. The block was submerged in deionized water inside a transparent chamber. A high-speed camera (Phantom VEO 710 L) was mounted directly outside the chamber and a light source was placed on the opposite side.

Datasets of two types of experiments were generated; transient and steady state. The transient cases consisted of increasing the power until the CHF was reached then quickly turning off the heaters. Due to limitations of camera memory, these cases used sampling rates of 150 frames per second to capture in order to capture the onset of nucleate boiling to the CHF in a continuous video. These transient cases were used for training and testing the segmentation models since the frame rate was too low for tracking purposes. The steady-state tests describe running the experiment for extended periods at constant power.

These cases used a high sampling rate of 3,000 frames per second since the videos could be at a shorter duration and still capture relevant features.

In total, over 1,000 images were labeled for training and over 200 for testing. These images were randomly taken from different boiling experiment datasets shown in TABLE 1 using a Python script. The images were used for the training and testing of two separate instance segmentation models. For the two-class model, the polygon tool in the LabelMe[26] software was used to manually label all of the images. For this labeling, bubbles were outlined and categorized as either "attached" or "detached" based on if the bubble was connected to the heater surface or not. Then, the data was saved to the MS COCO format under JSON files to be used for developing the model. For the one-class model, the same data and outlines were used but both "attached" and "detached" bubble categories were replaced with "bubble". For both models, annotations for one entire experiment, Boiling-2, were withheld from the training sets in order to test the model's performance on data from different experiments not used in training. In summary, the models were trained on subsets of data from boiling test datasets described in TABLE I (excluding Boiling-2) and the models were tested using a subset of Boiling-2 and a different than training subset of Boiling-1.

**B. Machine Learning Models**

Two different types of models (i.e., instance segmentation and tracking) are paired to extract important information from the high-speed images. These features can be split into individual bubble features, global features, and dynamic image features. The individual bubble features are based on a single bubble. The global features refer to the characteristics of all bubbles in single frames. The dynamic features are based on image sequences and utilize the results of the tracking model with the segmentation model to represent full-frame temporal data. FIG. 1 shows the overall model architecture. Image sequences are passed through the segmentation model and the tracking model for extracting masks of bubbles and labeling the same bubble in consecutive frames. Then, this information is used to extract the different bubble features.

TABLE I. Pool boiling datasets used for model training and testing.

| Dataset ID | Type | Surface | Frame Rate (fps) | Used in model training? | Source |
|---|---|---|---|---|---|
| Boiling-1 | Steady-State | Polished Cu | 3000 | Yes | Ref [24] |
| Boiling-2 | Steady-State | Cu microchannels | 3000 | No | Ref [25] |
| Boiling-3 | Steady-State | pH 0 copper foam | 3000 | Yes | Ref [24] |
| Boiling-4 | Steady-State | pH 10 copper foam | 3000 | Yes | Ref [24] |
| Boiling-5 | Transient | Polished Cu | 150 | Yes | Present work |
| Boiling-6 | Transient | Polished Cu | 150 | Yes | Present work |

For the framework, an instance segmentation model is utilized. In general, there are two types of segmentation models; semantic and instance. Semantic segmentation is used to classify each pixel of an image as a specific class (i.e., bubble or background). Instance segmentation on the other hand identifies objects and assigns a label to each object. Instance segmentation is used to distinguish between multiple objects that belong to the same class. For the work presented here, a pretrained instance segmentation model, Mask R-CNN [27], was used from the Detectron2 [28] GitHub repository. Mask R-CNN is an extension of the Faster R-CNN object detection. It was developed to address the task of instance segmentation, i.e. detect objects while precisely delineating their boundaries. It is a convolutional neural network (CNN). In general, a CNN works by establishing filters with random weights. Then, through the training process, data is passed through the model and these weights are continuously adjusted through iterations to minimize a designated loss function (objective function). The Mask R-CNN model is made of several CNN components. First, an input image is passed through a CNN for extracting features. Then, this output is passed through a region proposal network for generating predictions of the bounding boxes and identifying objects. The region proposal network outputs and initial feature extractions are then passed through a region of interest align layer in order to align features. Then, this is passed through another CNN to output the individual masks for the detected objects. It is also passed through layers to obtain both the classes of each object and the coordinates of bounding boxes for each object. In summary, the Mask R-CNN model takes image inputs and produces three separate outputs; individual masks of the objects, bounding boxes, and class labels for each object. The boiling image training datasets whose preparation was described in section A were used for finetuning the model. Two different segmentation models were trained separately using the manually labeled datasets. The first model had one class (i.e., bubble). This segmentation model just identified all the bubbles in the image. The next model had two classes in an effort to distinguish between bubbles that were attached to the heater and bubbles that were departed. The output of both of these models consisted of bounding boxes for each bubble, the predicted class, confidence values for each prediction, and binary masks for each identified bubble.

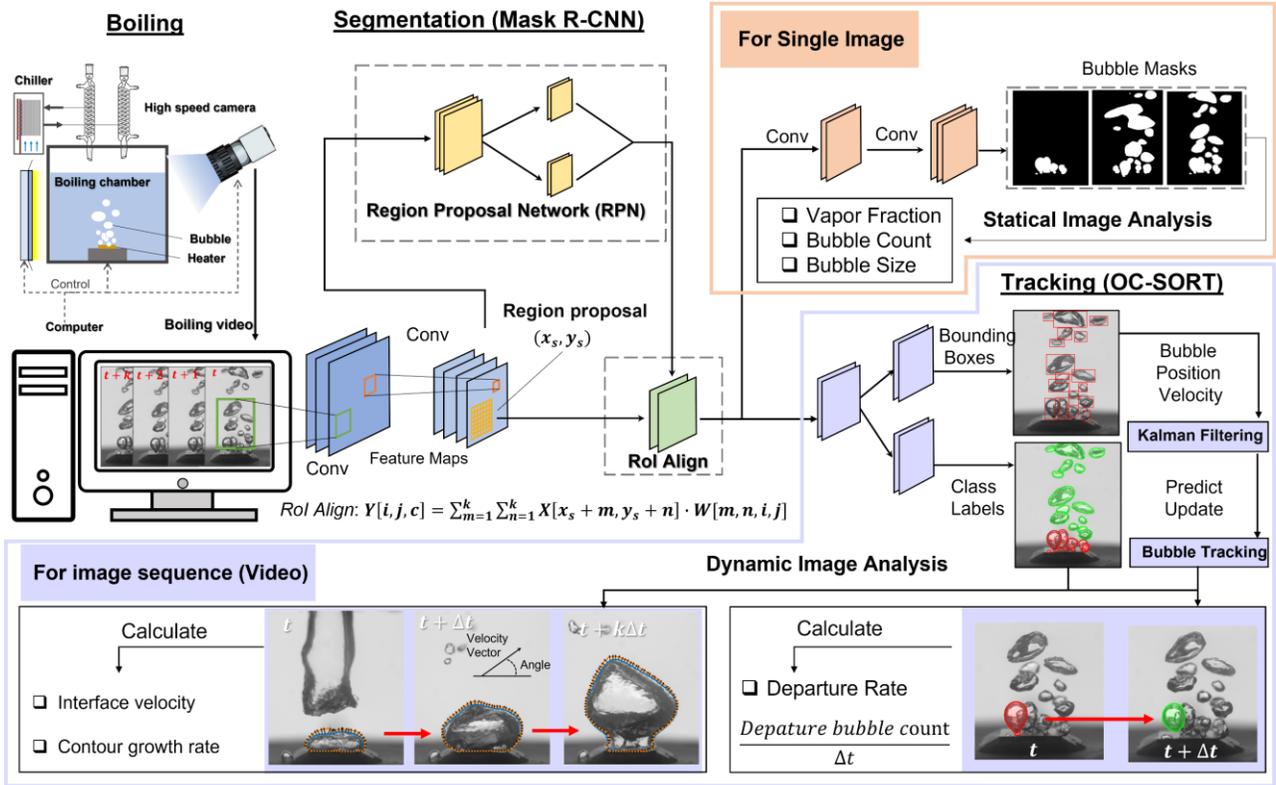

FIG. 1. Illustration of the BubbleID framework consisting of instance segmentation model (Mask R-CNN) and tracking model (OC-SORT) for obtaining individual bubble, global, and dynamic features.

The segmentation model was coupled with a tracking model to enable assigning matching IDs to the same bubble in consecutive frames. Tracking is made difficult due to several factors such as occlusion, noise, motion blur, deformation, etc. Simple online and real-time tracking (SORT) [29] has been the base of several new tracking models. SORT works by first using an object detection model to identify the location of the objects in each frame. Then, the Kalman filter is used to predict the next location of said object. The Kalman filter is a powerful algorithm used in many tracking applications and are used for predicting the state of an object. Some common issues with SORT are insufficient tracking robustness with non-linear motion and no observations for updating posterior. Also, there is a tradeoff between high framerate and performance. A higher frame rate is desired for the linear frame approximation but with a higher frame rate the noise object velocity variance is high. Noise in the velocity will accumulate into the position estimate. Also, the noise of the state estimate accumulates when there are no observations. Although SORT is a good baseline tracking model it has room for improving errors currently faced. Models such as DeepSORT [30] and OC-SORT [31] have been developed to improve issues seen by SORT. OC-SORT was a model proposed to improve performance with non-linear motion and occlusion and is the model used in this work. It includes a module, named observation-centric re-update, which uses object state observations to reduce accumulated error during tracks being lost, and an observation-centric momentum module for incorporating the direction consistency of tracks in the cost matrix for association.

## C. Bubble Dynamic Analysis Methodology

An important part of the proposed models is extracting physical meaning and features from the model outputs. Several different features can be extracted and computed from these types of machine learning models. These features are divided into individual bubble features, static global features, and dynamic features. The following describes how all the features are defined and how they are obtained from the machine learning segmentation and tracking models. For individual bubbles, many characteristics can be extracted. As a representation of bubble size, approximate bubble diameter was reported. This is defined as the diameter of a circle with the same area as the mask of the bubble and its formula is given below, where N is the number of pixels the bubble occupies based on the mask and $\alpha$ is a scale value that specifies the number of pixels in one *cm*.

$$Bubble\ diameter\ =\ \sqrt{N\frac{4}{\pi\alpha^2}} \qquad (1)$$

Another result reported was the bubble interface morphology which describes the location of the interface of a single bubble in a particular frame. This was achieved by taking the bubble masks produced from the Mask R-CNN model and using OpenCV's findcountours function [32]. The interface velocity of a bubble was presented as a new feature to obtain through segmentation and tracking. This describes how the bubble is expanding. To achieve these velocity vectors along the bubble interface first a single bubble was tracked through multiple frames. Then, an initial frame was taken with that bubble. Identically to the interface morphology extraction, the outline of the bubble was found by using OpenCV's findcountours function on a mask of that bubble and choosing the largest contour. This function provides coordinates of the contour of the mask. Next, this same process was used on the same bubble but 5 frames later as identified through the tracking model. Next, the contour of the bubble at the initial frame was compared to the contour at a future frame. Using cKDTree from the Scipy [33] library, the closest point on the second contour was matched to each point on the first contour. To convert these vectors to velocity vectors, the distance in pixels was converted to physical distance by using the 1cm wide heating surface as a reference. They were then divided by the time between the two frames.

The global data refers to features that are generated using all the bubbles in a single frame. One of the simplest examples of this is identifying **bubble count**. This is just the number of bubbles in each frame. This is achieved by summing the number of instances identified by the segmentation model in each frame. Vapor fraction describes the ratio between liquid and gas. Utilizing the two segmentation models, three representatives of the vapor fraction were obtained. The first was obtained by taking the number of pixels containing a bubble in the image and dividing it by the total number of pixels in the image. The second was obtained by taking the number of pixels containing bubbles classified as attached and dividing by the total number of pixels in the image. The third vapor fraction analysis reported the vapor fraction of a single bubble over time. This was achieved by using the tracking model to identify the same bubble from frame to frame and for each frame dividing the number of pixels that bubble occupied by the total number of pixels in the image.

Dynamic features are extracted over the time domain. The departure rate describes the frequency at which bubbles departure from the heater surface. The two-class segmentation model was used for determining the departure rate. Using the tracking model, each bubble was tracked through the video. Then, the classification results for each bubble through time were used. A departure event was defined when the status of a bubble was changed from attached to detached. To get the departure rate, this count of departed bubbles was divided by the duration of the clip.

## III. RESULTS AND DISCUSSION

### A. Machine Learning Model performance

To present the benefits of the proposed bubble dynamics analysis network, boiling experiments are conducted. Five metrics based on average precision (AP) [34–36] are used to verify its performance. Computing AP essentially involves calculating the area under the precision-recall curve. This is done by summing up the products of precision and the change in recall at each threshold, as given in equation (4), where precision and recall are calculated by equations (2) and (3). Precision is used to measure the accuracy of the model's positive predictions, while Recall is used to measure the model's ability to capture true positives. In these equations, $TP$ (True Positive) refers to the cases where the model correctly identifies a positive instance as positive. $FP$ (False Positive) refers to the cases where the model incorrectly identifies a negative instance as positive. $FN$ (False Negative) refers to the cases where the model incorrectly identifies a positive instance as negative.

$$Precision = \frac{TP}{TP + FP} \tag{2}$$

$$Recall = \frac{TP}{TP + FN} \tag{3}$$

$$AP = \sum_{n}(Recall_n - Recall_{n-1}) \times Precision_n \tag{4}$$

AP50 is the average precision calculated using an IoU (Intersection over Union) threshold of 0.5 when computing AP. IoU is a metric used to measure the overlap between the detected object and the ground truth, where an IoU of 0.5 means that a detection is considered correct if its overlap with the ground truth exceeds 50%. An IoU of 1 is a perfect prediction while 0 means no overlap. Compared to AP50, AP75 employs a higher IoU threshold, thus requiring a greater overlap between the detection and the ground truth. Attached AP and Detached AP are used to evaluate the detection performance of the model for bubbles attached to and detached from the heating surface, respectively. With these five metrics, the effectiveness of the proposed model can be comprehensively evaluated.

**TABLE II.** The performance of proposed machine learning model for bubble object detection and segmentation.

| Boiling Test | AP | AP50 | AP75 | Attached AP | Detached AP |
|---|---|---|---|---|---|
| Boiling-1 | 50.825 | 74.892 | 57.668 | 63.186 | 38.465 |
| Boiling-2 | 43.557 | 72.706 | 44.828 | 48.952 | 38.162 |

The evaluation results are provided in TABLE II**.** In Boiling-1, the model exhibited a high AP of 50.825%, reaching 74.892% in AP50 (IoU=0.5), indicating its good detection capability under a relatively lenient IoU threshold. Additionally, the AP75 (IoU=0.75) result of 57.668% suggests the model's ability to accurately segment bubbles under stricter conditions. Concerning the identification of different types of bubbles, the average precision for attached bubbles was 63.186%, whereas for detached bubbles, it was slightly lower at 38.465%, reflecting the greater difficulty in recognizing and tracking detached bubbles in complex backgrounds. The results of the Boiling-2 were lower, with an overall AP of 43.557%, AP50 of 72.706%, and AP75 of 44.828%. This disparity can be attributed to the fact that the model was trained without any images from the Boiling-2 dataset. This boiling was completely used as a test set. Therefore, even without including this boiling image in the training set, the model still achieved high accuracy in AP50, further demonstrating its generalization ability. The average precisions for attached and detached bubbles were 48.952% and 38.162%, respectively.

### B. Bubble dynamic analysis

Based on the segmentation and tracking result, the statistic and dynamic analysis of bubbles can be made. FIG. 2 detailing individual bubble information in FIG. 2A, global (spatially-averaged) information in FIG. 2B, and temporal-spatial (dynamic) information in FIG 2C. In FIG 2A, the detection, segmentation, and classification results of one frame taken from Boiling-1 at a heat flux of 13.97 W/cm$^2$ are provided. Using the proposed model, all the bubbles can be classified in to categories; attached and departure. They are marked in red and green in the left most image and marked as 0 and 1 in the right most table respectively. From this figure, it can be seen that the model does well segmenting and classifying the bubbles. By performing static analysis on the image, we can obtain diameter, interface morphology, and interface velocity of individual bubble. The distribution of diameter is provided in a histogram in the middle of FIG. 2A which illustrates that the frequency of bubbles within the diameter range of (0~5) increases initially and then decreases, reaching a maximum frequency at around *2mm*. In the rightmost table of FIG. 2A, we provide some parameters of four bubbles including some attached to the heating surface and some departed. As for the interface morphology and velocity of bubbles, we will elaborate on them in the subsequent discussion. Global information in FIG. 2B refers to information utilizing all the bubbles in a single frame, this includes bubble count and vapor fraction. The vapor fraction equation is given below

$$\varphi = \frac{V_{vapor}}{V_{total}} \tag{5}$$

Here we provide the bubble count at four different time points, along with the attached vapor fraction and total vapor fraction, as shown in FIG 2B. The ability of the model to distinguish between attached and departed bubbles enables the automatic acquisition of attached vapor fraction and total vapor fraction. The motivation behind this distinction of vapor fraction is that a bubble once detached plays less of a role in the heat transfer compared to the attached ones. The framework also enables the calculation of bubble departure rates. As shown in FIG 2C, the trend of bubble departure rate with heat flux is depicted for two heating surfaces, namely Plain Cu and Microchannel. It can be observed that they exhibit an inverse-proportional relationship. It also can be seen that both experiments with different surfaces show similar rates at around the same heat fluxes. This is an interesting observation because the two experiments had different critical heat fluxes; 102 W/cm$^2$ for the Plain Cu surface and 203 W/cm$^2$ for the Microchannel surface.

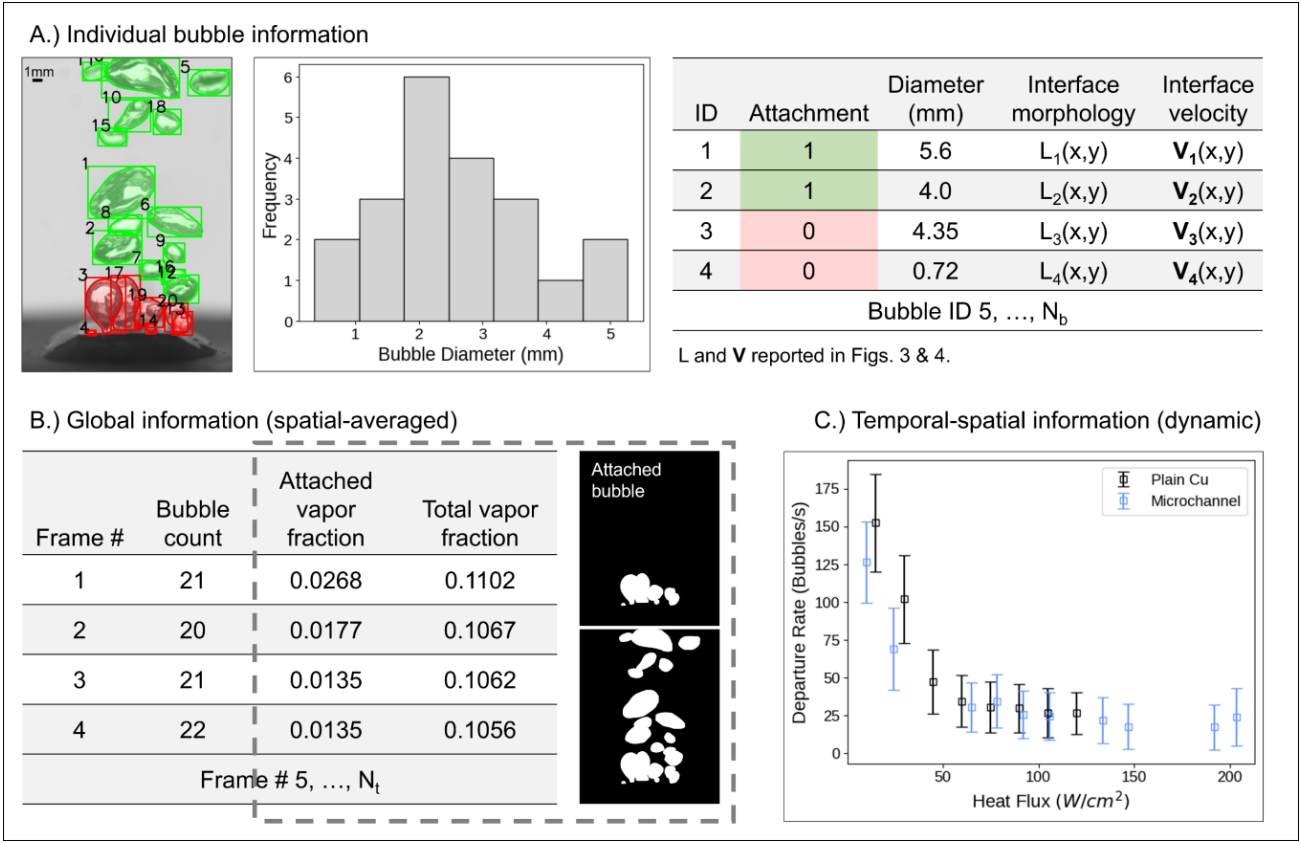

FIG. 2. BubbleID-extracted features showing (a) individual bubble features including bubble ID, diameter, attachment status, and interface morphologies of each individual bubble, (b) spatial-averaged information including bubble count, attached vapor fraction, and total vapor fraction of each frame, and (c) dynamic features including the bubble departure rate. Example features in (a) and (b) are from Boiling-1 at a heat flux of 13.97 W/cm$^2$ and data for (c) is from Boiling-1 and Boiling-2.

Through the proposed method, the dynamic growth characteristics of bubbles can be measured. As shown in FIG. 3a, the velocity direction of the bubble's interface at different positions is illustrated. With the instance segmentation of the bubble, obtaining the bubble's edge becomes feasible. Based on the acquired edges and tracking labels, a wealth of dynamic growth information can be quickly obtained. In FIG. 3b, we provide a reference direction for the bubble's interface velocity, where a vector on the interface is either pointing inside the bubble or outside and those pointing outside are positive. FIG. 3c defines the description method for positions on the bubble's interface. In this study, the position of a velocity vector is described in terms of relative perimeter. For each bubble, 0 is defined in the middle of the bottom of the bubble and then moves counter-clockwise around the bubble. FIG. 3d displays the dynamic variation of the bubble's interface velocity over time of an example case from Boiling-1 at a heat flux of 102.63 W/cm$^2$. This plot has a Gaussian filter applied to smooth the graph and highlight changes. The color bar only shows a range from -30 to 30 cm/s to highlight the different velocities. The vertical axis here is based on the annotations in FIG. 3c. Based on FIG. 3, it can be observed that the interface velocity around 1/2 of the bubble's perimeter is the highest, and with the increase of time, the interface velocity also increases. The range of bubble interfaces with positive velocities gradually increases until the bubble detaches from the heating surface. Similarly, the velocity of the interface near the heating surface changes from an approximately zero magnitude to a large negative magnitude when the bubble detaches.

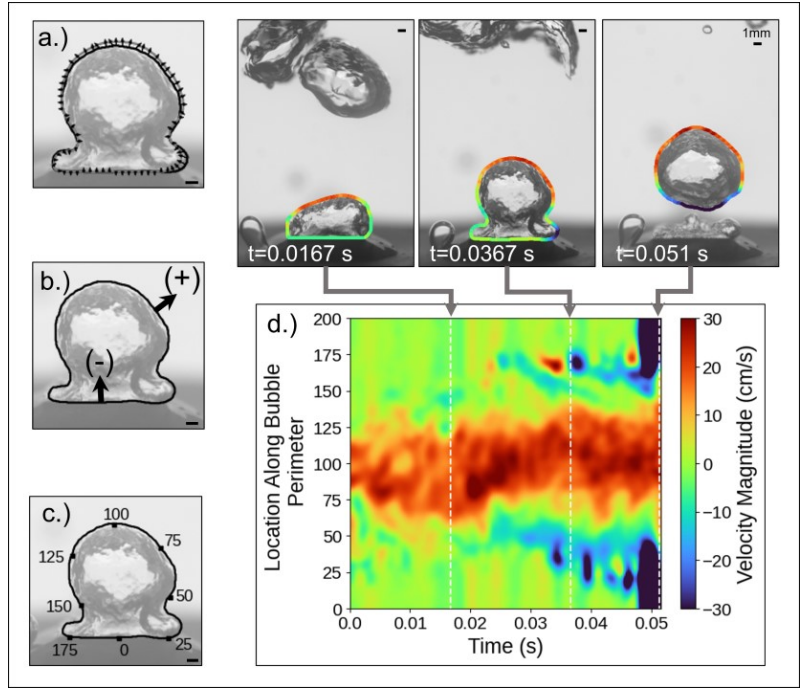

FIG. 3. Bubble interface dynamics analysis showing (a) bubble interface contour and velocity vectors, (b) definition of velocity signs (vectors pointing outside of bubble are defined as positive while vectors pointing inside bubble are defined as negative), (c) definition of locations on the bubble perimeter which is used as axis in (d), and (d) interface velocity profiles over time with representative images. This is an example case from Boiling-1 at a heat flux of 102.63 W/cm$^2$.

To further demonstrate the usage of bubble dynamic analysis, a comparative visualization of bubble behavior and dynamics before and after the critical heat flux (CHF) in a boiling process is provided in FIG. 4. These new example cases are from Boiling-1 at a heat flux of 93.92 W/cm$^2$ (pre-CHF) and Boiling-1 at a heat flux of 102.63 W/cm$^2$ (post-CHF). For each situation, the dynamic changes in the maximum velocity magnitude of a bubble interface over time, the velocity magnitude along the bubble perimeter over time, and velocity magnitude distribution pattern over time are provided. Maximum velocity magnitude is defined as the maximum absolute value velocity magnitude for the bubble in a single frame. Bubble vapor fraction is defined as the ratio of the bubble of interest's area in the image over the entire area of the image. These figures present a marked difference in bubble dynamics before and after the CHF is reached. Before CHF, the increasing curve of bubble maximum velocity magnitude is more stable compared with that of post-CHF which reveals more significant fluctuations. More notably, the peak value of bubble maximum velocity magnitude in pre-CHF when bubble departure is almost triple than that of post-CHF. The velocity magnitude along the bubble perimeter over time also shows different increasing patterns. During the rising period, both cases exhibit growth curves resembling concave functions with different curvatures. However, the inflection point of the bubble vapor fraction at pre-CHF shows a smoother decline, while the transition post-CHF is more abrupt. From the velocity magnitude distribution spectrum and the actual bubble growth-departure images, it's easier to get an intuitive difference between the two cases. In the pre-CHF state, the morphology of bubbles undergoes more pronounced changes over time, whereas in the post-CHF state, bubble growth becomes more stable and predictable. Therefore, it can be summarized that employing the proposed bubble segmentation-tracking model and interface velocity has great potential for identifying or distinguishing the pre/post-CHF state.

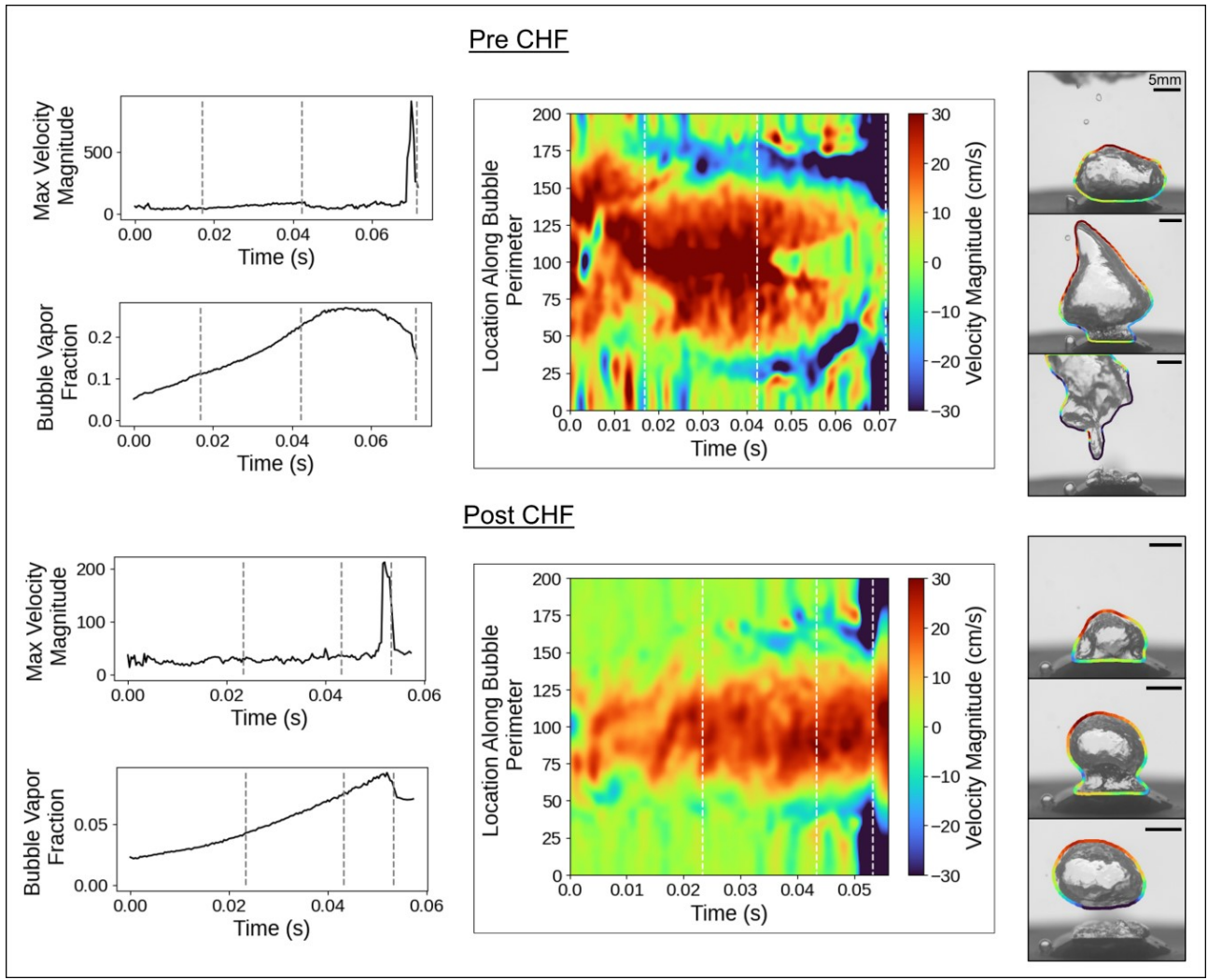

FIG. 4. Comparison between bubble dynamics before CHF (top panel) and after CHF (bottom panel) showing the velocity magnitude versus time, bubble vapor fraction versus time, and velocity profiles with representative bubble images. The pre-CHF test case is from Boiling-1 at a heat flux of 93.92 W/cm$^2$ and the post-CHF test case is from Boiling-1 at a heat flux of 102.63 W/cm$^2$.

## IV. CONCLUSION

Based on the experiment results, the following conclusions can be drawn.

1) The proposed machine learning framework is used in boiling bubble detection, segmentation, classification, and tracking.

2) From individual bubble analysis, the bubble diameter and distribution per frame can be found. The global analysis allows for vapor fraction calculation of both only bubbles attached to the heater surface and all of the bubbles in a single frame. Dynamic analysis is also made by determining the departure rate through the use of the class labels.

3) The proposed metric, interface velocity, is used in bubble dynamic analysis. Using this metric, the bubble growth-departure features can be comprehensively evaluated. Meanwhile, the experiment result suggests that the interface velocity of bubbles in different locations shows a different changing pattern. Furthermore, qualitative differences in dynamic features between pre-CHF and post-CHF are observed.

Overall, this work, in particular the interface velocity vectors, has potential for utilization in verifying some boiling CFD models. Future work will include further increasing the segmentation and tracking precision and improving the generalizability of the models. Meanwhile, a deeper analysis of bubble dynamics using machine learning and physical modeling based on the extracted features should also be carried out for a more fundamental understanding of bubble dynamics.

**ACKNOWLEDGMENTS**

This work was supported by the National Science Foundation (NSF) Grant No. CBET-2323022. The authors would like to thank Professor Ashif Iquebal and Ridwan Olabiyi at Arizona State University for their valuable discussion about this work. The authors are grateful to Amanda Williams and Ethan Weems at the University of Arkansas for their assistance in annotating boiling images.

This paper has been submitted to the Journal of Applied Physics.

**DATA AVAILABILITY**

The data that support the findings of this study are available from the corresponding author upon reasonable request.